\documentstyle[12pt,aaspp, flushrt]{article}
\def\ls{{_<\atop^{\sim}}}
\def\gs{{_>\atop^{\sim}}}
\begin{document}

\title{X-ray Spectral Survey of WGACAT Quasars, I:\\
Spectral Evolution \& Low Energy Cut-offs}

\author{Fabrizio Fiore$^{1,2,3}$, Martin Elvis$^1$, Paolo Giommi$^3$, \\
Paolo Padovani$^4$}

\affil{$^1$ Harvard-Smithsonian Center for Astrophysics\\
60 Garden St, Cambridge MA 02138}

\affil{$^2$Osservatorio Astronomico di Roma, Monteporzio (Rm), Italy}

\affil{$^3$BeppoSAX Science Data Center, Roma, Italy}

\affil{$^4$Dipartimento di Fisica, II Universit\`a di Roma}

\author{\tt version: 1pm June 24 1997}

\begin{abstract}

We have used the WGA catalog of ROSAT PSPC X-ray sources to study the
X-ray spectrum of about 500 quasars in the redshift interval 0.1--4.1,
detected with a signal to noise better than 7. We have parameterized
the PSPC spectrum in terms of two `effective energy spectral indices',
$\alpha_S$ (0.1-0.8 keV), and $\alpha_H$ (0.4-2.4 keV), which allows
for the different Galactic $N_H$ along the quasars line of sight.  We
have used these data to explore the questions raised by the initial
PSPC high redshift quasar studies, and in particular the occurrence of
low X-ray energy cut-offs in high redshift radio-loud quasars.  We
have also studied the emission spectra of a large sample of radio-loud
and radio-quiet quasars and studied their differences.

We find that low energy X-ray cut-offs are more commonly (and perhaps
exclusively) found in radio-loud quasars.  Therefore the low energy
X-ray cut-offs are physically associated with the quasars, and not
with intervening systems, since those would affect radio-quiet and
radio-loud equally.  We suggest that photoelectric absorption is a
likely origin of the these cut-offs.

The number of `cut-offs' in radio-loud quasars significantly increases
with redshift, rather than with luminosity.  A partial correlation
analysis confirms that $\alpha_S$ is truly anti-correlated with
redshift at the $99.9\%$ confidence level, indicating evolution with
cosmic epoch, and not a luminosity effect.  Conversely, for $\alpha_H$
the observed anti-correlation with redshift is mostly due to a strong
dependence on luminosity.

We find marginal evidence for a flattening of $\alpha_H$ (P=4.5 \%)
going from z$<$1 to z=2, in radio-quiet quasars, in agreement with
previous studies.  On the other hand, radio-loud quasars at z$<$2.2
show a `concave' spectrum ($\alpha_H < \alpha_S$ by $\sim$0.2). This
new result is consistent with the widespread suggestion that the X-ray
spectrum of radio-loud quasars may be due to an additional component
above that seen in radio-quiet quasars.  However, it might also imply
different processes at work in radio-loud and radio-quiet sources.  At
z$\gs2$ the average soft and hard indices are similar and both
significantly smaller than at lower redshifts. This can be due to the
soft component of radio-loud quasars being completely shifted out of
the PSPC band at z$>2$.

\end{abstract}


\section{Introduction}

The ability of ROSAT to study fainter X-ray sources than ever
before opened up the range of quasars that could be reached to
span virtually the whole of their properties.  High redshift
objects (up to z$\approx4$) became accessible in X-rays, and their
spectrum was measured for the first time between 0.4 and 10 keV
(in the quasar frame, Elvis et al 1994a). Unexpected low energy
cut-offs, far larger than those expected due to absorption by our
galaxy, were detected in several high-z radio-loud quasars
(Wilkes et al. 1993, Elvis et al. 1994a). The obvious possibility
was that these were caused by photoelectric absorption, either
along the line of sight, or at the quasar. If the absorber were
at the quasar, then the material could be nuclear, as in low
redshift low luminosity objects (e.g. Elvis \& Lawrence, 1985, and
Elvis, Mathur \& Wilkes 1995)
or could be on the larger
scale of the host galaxy or proto-galaxy. A tentative link with
the highly compact `GigaHertz Peaked' (GPS) radio sources
suggested the latter and hence that X-ray astronomy offered a new
probe of early galaxy conditions.

Targeted ROSAT studies of high redshift quasar X-ray spectra are,
however, limited to a dozen or so objects. Within the more than
3000 ROSAT PSPC pointings (covering about 10 \% of the sky) lie
more than 50,000 sources (White, Giommi \& Angelini 1995, Voges
et al 1994).  Out of these, a sample of several hundred quasar
X-ray spectra can be readily compiled for the first time, thanks
to the release of the two source catalogs, WGACAT (White, Giommi
\& Angelini 1995) and ROSATSRC (Voges et al 1994). We have used
these data to explore the questions raised by the initial PSPC
high redshift quasar studies. As an additional benefit of this
program we are able to study the emission spectra of a large
sample of radio-loud and radio-quiet quasars and study their
differences.

We investigate the connection between low energy X-ray cut-offs
and the radio and optical spectra of these quasars in a companion
paper (Elvis et al.  1997, Paper II).  A later paper (Nicastro et
al. 1997) will discuss a sub-sample of quasars selected to have
the typical colors produced by absorption features imprinted on
the X-ray spectrum by an ionized absorber along the line of
sight.

We use a Friedman cosmology with H$_0$=50~km~s$^{-1}$~Mpc$^{-1}$
and $\Omega = 0$ throughout this paper.


\section{The Sample}

Quasars were selected by cross-correlating the first revision of the
WGA catalog (White, Giommi \& Angelini 1995) with a variety of optical
and radio catalogs, including the Veron-Cetty \& Veron (1993) and
Hewitt \& Burbidge (1993) quasar catalogs, and the 1 Jy (Stickel et
al. 1994) and S4 (Stickel \& K\"uhr 1994) radio catalogs. Uncertain
classifications and border line objects have been also checked in the
Nasa Extragalactic Database (NED).
\footnote{The NASA/IPAC Extragalactic Database (NED) is
operated by the Jet Propulsion Laboratory, California Institute
of Technology, under contract with the National Aeronautics and
Space Administration.}
All the objects selected have
optical spectra dominated by nonstellar emission and all show
broad emission lines.


The maximum radius adopted for identifying cross-correlation
candidates was one arcminute. The resulting distribution of
X-ray/optical offsets is shown in Figure 1.  The mean offset is
$\simeq 18$ arcsec, while the median one is $\simeq 13$ arcsec,
in agreement with the estimated errors on the positions of the
WGA sources (White et al. 1995).  Potential mis-identifications
through chance co-incidences were addressed by shifting the X-ray
positions by various amounts several times, and repeating the
cross-correlations. Using these randomized X-ray positions
establishes the chance co-incidence rate to be small. The number
of spurious X-ray/optical associations is at maximum 2, i.e.
less than 0.5\% of the whole sample (see below).


We excluded from the sample:
\begin{enumerate}
\item
observations of sources with a quality flag in the WGA catalog
$<5$ (corresponding to problematic detections);
\item
observations of sources with a signal to noise ratio in the
0.1-2.4 keV energy band less than 7 (to ensure reasonable X-ray color
determinations);
\item
observations of quasars located at an off-axis angle larger than
45 arcmin, (to avoid large systematic errors due to the
uncertainties in the PSPC calibration near the edge of the
field of view);
\item
fields with a Galactic $N_H$ along the line of sight higher than
$6\times10^{20}$ cm$^{-2}$ (to ensure good low energy signal to
noise);
\item
quasars with z$<0.1$ (to eliminate the low luminosity Seyfert
galaxies in which absorption is common, e.g. Lawrence \& Elvis
1982).
\end{enumerate}

We are then left with 453 quasars for which fluxes in three
bands, and so two X-ray colors (``soft'' and ``hard''), can be
derived. This is the largest sample of quasars for which
homogeneous X-ray spectral information is available.  We used
only one observation for each quasar.  When more than one
observation was available for a quasar we chose the one with the
highest signal to noise ratio.

Of these quasars 202 have a radio measurement in the literature,
and 167 of those are radio-loud according to the usually adopted
definition $R_L=log(f_{\rm r}/f_{\rm B}) > 1$, with $f_{\rm r}$
radio flux at 5 GHz and $f_{\rm B}$ the B-band flux (Wilkes \&
Elvis 1987; Kellermann et al. 1989; Stocke et al. 1992). This
translates into a (rest-frame) value of the radio to optical
spectral index, $\alpha_{\rm ro} > 0.19$.  The sample includes 87
Flat Spectrum Radio Quasars ($\alpha_{\rm r} < 0.5 $), the
majority of which are discussed by Padovani et al (1997).  Steep
Spectrum Radio Quasars account for 62 sources. Note that this
flat/steep classification is based mostly on the radio spectrum
at only 2 or 3 frequencies.  Quasars with complex radio spectra
(e.g. GigaHertz Peaked Spectrum (GPS) Sources) could appear in
either class. This classification also does not distinguish
compact steep spectrum (CSS) radio sources from extended ones.
For 18 radio-loud quasars we could find radio measurements at one
frequency only, and therefore their radio spectral index is
unknown. We assumed $\alpha_R=0.5$ in calculating
$\alpha_{ro}$.  Most of the quasars without radio measurement
are likely to be radio-quiet (e.g. Ciliegi et al., 1995) so we
include them in the radio-quiet sample, making for 286
radio-quiet quasars in the sample.

In the WGACAT there are another 35 quasars with Galactic
$N_H>6\times10^{20}$ but passing all other above points, for
which we derive a ``hard'' color only. Of these 28 are
radio-loud.

The redshift distributions for radio-quiet quasars and radio-loud
quasars is shown in figure 2.  At z$\ls1$ the number of
radio-quiet objects is higher than that of radio-loud, due to the
higher volume density of radio-quiet quasars. By z$\gs2$ however,
the number of radio-loud objects is higher than that of the
radio-quiet objects, due to the higher $L_X/L_{opt}$ of
radio-loud quasars.

\section{Effective X-ray Spectral Indices}

The WGA catalog provides raw count rates in three energy bands:
0.1-0.4 keV (soft band, S), 0.40-0.86 keV (mid band, M) and 0.87-2 keV
(hard band, H) for each entry.  The count rates can be combined to
form a `softness ratio' (SR=S/M) and a `hardness ratio' (HR=H/M).  To
allow for the effect of the varying Galactic absorption from source to
source these softness and hardness ratios can be converted to
`effective spectral energy indices', $\alpha_S$ (0.1-0.8 keV), and
$\alpha_H$ (0.4-2.4 keV), assuming the Galactic $N_H$ as derived from
21 cm measurements (Stark et al 1992, Shafer et al, private
communication, Heiles \& Cleary 1979),``Wisconsin'' cross-sections
(Morrison \& McCammon (1983) and solar abundances (Anders \& Ebihara
1982), and using the PSPC calibration.

\subsection{Correction for the instrumental point spread function}

In converting the three count rates into effective spectral indices,
corrections must be applied for the instrumental point spread function
(PSF), to take into account the different fraction of counts lost
outside the detection region in each energy band.  The total PSPC PSF
includes two main components: the mirror PSF and the PSPC detector PSF
(Hasinger et al 1992a). The mirror PSF is energy independent, and
dominates the total PSF at off-axis angles greater than about 20
arcmin. At these large off-axis angles a small detection region loses
flux but does not significantly alter the spectral shape.

The detector PSF is, instead, highly energy dependent, and dominates
the total PSF at small off-axis angles. Using a small detection region
for an off-axis angle $<20$ arcmin can strongly alter the spectral
shape, predominantly at low energies. (The detector FWHM$\propto
E^{-0.5})$. The method used in the WGA catalog to estimate the source
count rates uses the counts detected in a box whose size optimizes the
signal to noise ratio for each source. For relatively weak sources
near the field center, this optimum box size is small, and the number
of soft photons lost is greater than that of medium or hard energy
photons, giving rise to an artificial reduction of the spectrum at low
energies, which in turn can lead to an underestimation of $\alpha_S$
of up to 0.2-0.3.

We corrected the count rates in the three bands for both mirror and
PSPC PSF effects. We used the analytical approximations for the energy
and off-axis angle dependence of the PSF provided by Hasinger et al
(1992b).

This method of estimating spectral indices is similar to the
de-reddening procedure used in optical photometry, and is quite
robust, with typical systematic uncertainties on $\alpha_H$ smaller
than 0.1 (as shown, for example, by Padovani \& Giommi 1996 or Ciliegi
\& Maccacaro 1996). It is particularly suitable for handling large
samples of objects (Giommi et al., in preparation) e.g. for the
determination of the X-ray spectral index distributions.

However, a systematic uncertainty is present in the spectral indices
estimates, because the magnitude of the PSF correction depends on the
intrinsic source spectrum. To estimate the magnitude of this
uncertainty we calculated five series of $\alpha_S$ and $\alpha_H$ for
a grid of assumed source spectra: for the average value for high
Galactic latitude line of sight ($N_H= 3\times10^{20}$ cm$^{-2}$) we
used $\alpha_E=1$, $\alpha_E=1.5$ and $\alpha_E=2.5$; for the mean
value of energy spectral index ($\alpha_E$=1.5) we used low ($N_H=
10^{20}$ cm$^{-2}$) and high ($N_H= 10^{21}$ cm$^{-2}$) absorption
values.  We then calculated the differences between the $\alpha_S$ and
$\alpha_H$ calculated for each pair of $\alpha_E$ and $N_H$ and those
calculated for $\alpha_E=1.5$ and $N_H= 3\times10^{20}$ cm$^{-2}$.
The differences in $\alpha_H$ were always smaller than 0.05 and those
in $\alpha_S$ were always smaller than 0.15. We therefore use these
values as systematic uncertainties in the evaluation of these
parameters.


\subsection{PSPC background}

The WGACAT count rates are not background subtracted. Therefore,
another possible source of error in the evaluation of the PSPC
spectral indices is the PSPC background in the three WGACAT bands. The
PSPC background has been studied in great detail by Snowden et al
(1992, 1994, noncosmic background, and 1995, 1997 diffuse, cosmic
X-ray background) using ROSAT All Sky Survey (RASS) data and pointed
observation data.  The energy band which shows, by far, the highest
background at high Galactic latitude is the 0.1-0.28~keV range, which
spans most of the SOFT band defined in WGACAT (0.1-0.4~keV) We now
analyze the effect that a wrong or absent background subtraction in
the SOFT band can have on the soft energy index $\alpha_S$.

The cosmic background in the SOFT band is highly spatially variable,
with large regions of maxima of about 0.0015 counts~s$^{-1}$
arcmin$^{-2}$ at high Galactic latitudes ($|b|>30$) and minima of
0.0003 counts s$^{-1}$ arcmin$^{-2}$. Long-Term enhancements are
difficult to identify and model in pointed observations. They mainly
affect the low frequency part of the spectrum, E$<0.28$ keV, where
their contribution can even be as high as 0.001 counts s$^{-1}$
arcmin$^{-2}$. We adpot a conservative value for the total SOFT band
background of 0.003 counts~s$^{-1}$~arcmin$^{-2}$.

The typical box-side of the WGACAT extraction regions is 0.3-0.8
arcmin for off-axis angles smaller than 20 arcmin and 1-3 arcmin for
off-axis angles between 20 and 45 arcmin, where the lower limits apply
to the faintest sources.  Such extraction regions, together with the
above background rate would give a few counts in the low energy band
in typical exposures of 1000-10000 sec for sources detected in the
central 20 arcmin, and a few tens of counts for sources detected in
the outer PSPC region.

Let us assume the case of a faint (close to our limit of 7 in signal
to noise ratio) and strongly cutoff source, whose counts in the SOFT
band are a few, comparable with the background contribution, and 15-30
in the MEDIUM band.  In this case the upper limit on the slope between
the SOFT and MEDIUM energy bands differs from the slope we would
measure neglecting background subtraction by $\sim 0.5$.  This
compares with a typical statistical error in $\alpha_{S}$ of 0.5-0.7
for such faint sources and with a systematic uncertainty of 0.15 We
conclude that the small extraction regions used in the WGACAT means
that background is not a significant problem, even for the SOFT band
count rate of our faint sources.  As a result we did not subtract
background in the evaluation of the effective spectral indices
$\alpha_S$ and $\alpha_H$.

To investigate the robustnes of this assumption, we searched for
correlations between the soft and hard effective spectral indices and
(1) the size of the extraction region, (2) the off-axis angle, and (3)
the source count rate. Any correlations would be evidence that
neglecting the background was causing problems.  In no case did we
find significant correlations. We conclude that neglecting the
background subtraction does not strongly affect or bias our results.


There are other two sources of uncertainties in the spectral indices
estimates: first, quasars located near the PSPC rib structures can
suffer a preferential loss of soft photons. We inspected the original
ROSAT images for each source that appeared to have unusually low soft
band counts and rejected those that might have been so
affected. Second, the spectral indices estimates from hardness ratios
may be significantly different from the results of a proper spectral
fit to the full PHA spectrum, in presence of a curvature in the
intrinsic spectrum, because of the skewness induced in the broad
energy bands used to construct the hardness ratios.  For all these
reasons the effective spectral indices $\alpha_S$ and $\alpha_H$
should not be regarded as a measure of the true emission spectral
indices. They should rather be regarded as a rough estimation of the
``average'' soft and hard spectral shapes. They are the soft X-ray
analogs of (U-B), (B-V) colors, for which multiple physical
interpretations are possible. We will use $\alpha_S$ as an indicator
of a possible low energy cut-off.

\subsection{Effective Spectral Indices and Quasar Properties}

The radio-quiet and radio-loud quasars $\alpha_S$ and $\alpha_H$
are plotted against each other in figure 3a,b. In the
radio-loud plot flat spectrum radio sources are identified by
circles, and steep spectrum sources by squares. High redshift
quasars (z$>$2.2) are shown with filled symbols. The range of
$\alpha_S$ and $\alpha_H$ are large and so the different parts of
the diagram correspond to radically different spectral
shapes. These are illustrated with three-point spectra in figure
3. For the purpose of this paper we are most interested in those
for which a low energy cut-off is indicated (lower center of
figure 3).

To study these features more closely we divided the quasars into
four redshift bins: 0.1-0.5; 0.5-1; 1-2.2; and z$>2.2$. Table 1
gives the average spectral indices ($\alpha_S, \alpha_H$) and
their dispersions ($\sigma(\alpha_S), \sigma(\alpha_H)$) in these
four redshift bins for both radio-quiet and radio-loud quasars.
The typical statistical uncertainties on $\alpha_S$ and $\alpha_H$ are
$\pm$0.2, the systematic uncertainty is at most $\pm0.15$ for $\alpha_S$
and $\pm0.05$ for $\alpha_H$, so the measured dispersions are not strongly
affected by these uncertainties.

\begin{table}[ht]
\caption{\bf Quasars Average Spectral Indices
		\& Cut-off statistics}
\begin{centering}
\begin{tabular}{lrlllllcl}
\hline
z bin & N & $\alpha_S$& $\sigma(\alpha_S)$ & N &
$\alpha_H$&$\sigma(\alpha_H)$
& no. cut-off & fraction \\
\hline
\multicolumn{8}{l}{Radio-Quiet Quasars} \\
0.1--0.5 &136& 1.73& 0.48 & 141 & 1.62& 0.51 & 7& 0.051\\
0.5--1.0 & 66& 1.68& 0.36 & 67  & 1.67& 0.51 & 3& 0.045\\
1.0--2.2 & 72& 1.67& 0.38 & 72  & 1.51& 0.49 & 1& 0.014 \\
$>$2.2   & 12& 1.54& 0.26 & 13  & 1.32& 0.69 & 0& 0 \\
\multicolumn{8}{l}{Radio-Loud Quasars} \\
0.1--0.5 & 34& 1.49& 0.52 & 41  & 1.20& 0.63 & 1& 0.029\\
0.5--1.0 & 45& 1.30& 0.62 & 51  & 1.18& 0.48 & 2& 0.044\\
1.0--2.2 & 70& 1.37& 0.52 & 78  & 1.13& 0.51 & 4& 0.057\\
$>$2.2   & 18& 0.89& 0.71 & 25  & 0.89& 0.60 & 7& 0.389\\
\multicolumn{8}{l}{Flat Spectrum Radio Quasars} \\
0.1--0.5 & 9 & 1.38& 0.30 & 14  & 1.32& 0.38 & & \\
0.5--1.0 & 26& 1.24& 0.73 & 30  & 1.11& 0.51 & & \\
1.0--2.2 & 40& 1.35& 0.50 & 43  & 0.99& 0.46 & & \\
$>$2.2   & 12& 0.74& 0.75 & 17  & 0.78& 0.60 & & \\
$>$0.1   & 87& 1.24& 0.63 & 104 & 1.04& 0.51 & & \\
\multicolumn{8}{l}{Steep Spectrum Radio Quasars} \\
0.1--0.5 & 19& 1.52& 0.62 & 21  & 0.96& 0.65 & & \\
0.5--1.0 & 14& 1.29& 0.44 & 16  & 1.23& 0.39 & & \\
1.0--2.2 & 24& 1.40& 0.58 & 28  & 1.12& 0.46 & & \\
$>$2.2   & 5 & 1.15& 0.58 & 7   & 1.33& 0.60 & & \\
$>$0.1   & 62& 1.37& 0.56 & 72  & 1.14& 0.56 & & \\
\hline
\end{tabular}
\end{centering}
\end{table}

\subsubsection{Radio-Loud vs. Radio-Quiet}

The mean and dispersion of $\alpha_H$ and $\alpha_S$ for the
radio-quiet quasars are ($1.59, 0.52$) and ($1.69, 0.41$); those
of radio-loud quasars are ($1.13, 0.55$) and ($1.32, 0.59$). The
difference of $\sim$0.5 in $\alpha_H$ agrees with the
widespread finding that radio-loud quasars have flatter X-ray
spectra than radio-quiet quasars (e.g. Elvis \& Wilkes 1987, Laor et
al., 1994, Laor et al., 1997, Schartel et al., 1996a).

For radio-quiet
quasars the distribution of $\alpha_S$ is consistent with that of
$\alpha_H$ in all redshift bins. In the last redshift bin there
are only 12 quasars (see Table 1), and only three at z$>2.5$,
and therefore the test is not very
stringent for this redshift interval.
This uniformity suggests a single emission mechanism dominating
the whole ROSAT band (c.f. Laor et al., 1997), at least
for redshifts smaller than about 2.

Radio-loud quasars, instead, have $\alpha_H$ smaller than
$\alpha_S$ by $\sim$0.2 (i.e. a concave spectrum) in the low
redshift bins (z$<$2.2). The Kolmogorov-Smirnov probability of
$\alpha_S$ being drawn from the same distribution function as
$\alpha_H$, is 5.0 \% for redshifts 0.1$-$0.5, 22 \% for
redshifts 0.5$-$1.0 and 0.14 \% for redshifts 1.0$-$2.2. This new
result could be interpreted in terms of an additional component
in the spectrum of radio-loud quasars above that seen in
radio-quiet quasars (as suggested earlier by e.g. Wilkes \& Elvis
1987). However, it may also imply different processes at work in
radio-loud and radio-quiet quasars, (as also suggested by Laor et
al., 1997). To disentangle these two possibilities a careful
analysis of the optical to X-ray Spectral Energy Distribution of
WGACAT quasars is needed. This is beyond the scope of this
paper and will be addressed in a forthcoming paper.

We have computed the mean $\alpha_S$ and $\alpha_H$ for the two
samples of flat and steep radio spectrum quasars. These are
reported in Table 1.  Although both distributions of $\alpha_S$
and $\alpha_H$ for the steep and flat radio quasars are
consistent with being drawn from the same distribution function
(using a K-S test), the mean $\alpha_S$ of steep radio spectrum
quasars is steeper than that of flat radio spectrum quasars (at
the 90 \% confidence level).  The mean $\alpha_H$ of the two
samples of quasars is on the other hand very similar. Redshift
bins 0.1-0.5 and z$>2.2$ are populated by too small a number of
flat radio spectrum and steep radio spectrum quasars respectively
to allow a statistically significant comparison.  In the other
two redshift bins the distributions of $\alpha_H$ and $\alpha_S$
of the two samples are consistent with each other.

The dispersion in $\alpha_H$ and $\alpha_S$ is large for both
radio-quiet and radio-loud quasars.
The radio-loud dispersion in $\alpha_S$ is larger than that of
radio-quiet quasar at a confidence level of 96 \%.  Emission or
absorption mechanisms that produce more varied outputs seems to be
needed for radio-loud quasars.

{ }From figure 3 it appears that the number of radio-loud quasars
with $\alpha_S<0.5$ is much larger than the number of radio-quiet
with $\alpha_S<0.5$, especially at high z (solid symbols).  This
is the sense of the change in $\alpha_S$ to produce low energy
cut-offs, as would be produced by photoelectric absorption.

We compare the distribution of radio-loud and radio-quiet indices
about their respective mean $\alpha_S$ (figure 4).  By offsetting
from the mean of each group we remove the difference in the group
means discussed above. Figure 4 shows the broader
dispersion of $\alpha_S$ for radio-loud quasars. It also shows a
population of radio-loud quasars with smaller $\alpha_S$,
i.e. quasars that show low energy cut-offs. This tells us with
high confidence that the cut-offs are not due to intervening
material. Intervening material would not `know' whether a
background quasar was radio-loud or radio-quiet. The cause of the
cut-offs must be physically associated with the quasar in some
way.

\subsubsection{Dependence on Redshift}

To quantify the differences with redshift and study their
evolution we plot $\alpha_H$ and $\alpha_S$ versus redshift for
radio-quiet and radio-loud quasars in figures 5a,b and 6a,b
respectively. There is no strong evidence of evolution of either
$\alpha_H$ or $\alpha_S$ with redshift for radio-quiet quasars.
Again the small number of radio-quiet quasars with z$>$2 does not
allow us a strong conclusion to be drawn for high redshift
objects.  For the bins 0.5$<$z$<$1 and 1$<$z$<$2.2 the
probability that the two distributions of $\alpha_H$ are draw
from the same distribution function is marginally unacceptable
(P=4.5\%), in agreement with previous studies (Schartel et al
1996b).

On the other hand radio-loud quasars show strong changes in both
$\alpha_S$ and $\alpha_H$ for $z>2.2$, in the sense that both indices
are smaller than in lower redshift bins.  We ran a Kolmogorov-Smirnov
test between the distributions of spectral indices in each possible
pair formed with the four redshift bins.  The distributions of
$\alpha_S$ and $\alpha_H$ at z$<2.2$ are all consistent with each
other.  Only the $z>2.2$ bin shows significant differences from the
others, for both $\alpha_S$ and $\alpha_H$.  Table 2 gives the
percentage probability that the spectral index distributions of
$\alpha_S$ and $\alpha_H$ in two redshift bins are drawn from the same
distribution function.

\begin{table}[ht]
\caption{\bf Kolmogorov-Smirnov test for Radio Loud Quasars}
\begin{centering}
\begin{tabular}{ccccc}
\hline
         &          & $\alpha_S$ &        &         \\
\hline
         & 0.1--0.5 & 0.5--1.0 & 1.0--2.2 & $>$ 2.2 \\
$>$2.2   & 3.0      & 5.8        & 2.1    &  --   \\
\hline
         &          & $\alpha_H$ &        &         \\
\hline
$>$2.2   & 0.10     & 4.1        & 0.16   &  --  \\
\hline
\end{tabular}
\end{centering}
\end{table}

Some caution is required however. In flux limited samples
redshift and luminosity are often degenerate, so that it is hard to
distinguish the effects of one from the other. The apparent trend with
redshift discussed above may thus be induced by a correlation with
luminosity (figure 7a). We tested for this degeneracy by selecting a
sub-sample of only the high log($L_{opt}$) ($>$32) quasars. The correlation
between $\alpha_S$ and z for this sample is still very good (figure
7b, linear correlation coefficient $r=-0.46$, which, for 43 points,
corresponds to a probability of 99.8\%). On the other hand, the
correlation between $\alpha_H$ and z for the high luminosity, high
redshift quasar is not significant, $r=-0.17$ (probability of 72\%).

To better disentangle the luminosity/redshift dependence of $\alpha_S$
and $\alpha_H$  we also performed a partial correlation analysis
(e.g., Kendall \& Stuart 1979) on the whole radio-loud sample. While
$\alpha_S$ is anti-correlated with redshift even subtracting the effect of
the optical luminosity ($P \simeq 99.9\%$), in the case of $\alpha_H$
no correlation with redshift is left once the luminosity dependence is
subtracted out ($ \simeq 48\%$). On the other hand, excluding the
redshift dependence, we find that $\alpha_H$ is anti-correlated with
luminosity at the $99.2\%$ level.

We conclude that the flat $\alpha_S$, unlike $\alpha_H$, are truly
more common at high redshifts, and so are an evolutionary, or
cosmological, effect.

\subsubsection{Low Energy Cut-offs}

The high incidence of low energy cut-offs at high redshifts can
be made clearer by defining a sample of `candidate cut-off'
objects using $\alpha_S$ and then examining the fraction that
occur among both radio-loud and radio-quiet quasars as a function
of redshift.

The change in index of radio-loud quasars can be seen in figure
6b as due to a population of radio-loud quasars with a soft
spectral index significatively smaller than the
average and smaller than the hard spectral index.
(Although there are also a few radio-quiet quasars with
exceptionally flat $\alpha_S$, figure 5b). These small $\alpha_S$
quasars are `cut off' in their low energy spectrum compared with
an extrapolation of the hard spectrum.

We select `candidate' low energy cut-off quasars using three
criteria. (1) $\alpha_S<\alpha_H$;
(2) $\alpha_S<0.5$ selects all the high redshift radio-loud
candidates apparent in figure 6b. (3) Lower redshift radio-loud and
all radio-quiet candidates require a more careful selection since
their mean $\alpha_S$ are different. We use the criterion that
they have $\alpha_S$ smaller than the average by at least 0.75.
For radio loud quasars the average $\alpha_S$ at z$<2.2$ is 1.32
and therefore we selected quasars with $\alpha_S<0.58$ in this
redshift interval.  For radio-quiet quasars there is no evidence
for a change in the average $\alpha_S$ with z and therefore we
selected the quasars with $\alpha_S<0.94$ (the average $\alpha_S$
is 1.69).  These criteria select 10 radio-quiet and 14 radio-loud
`candidate' cut-off quasars. Table 3 gives the optical or radio
names, redshift and optical luminosity of these objects.

\begin{table}[ht]
\caption{\bf `Candidate' Cut-off Quasars}
\begin{centering}
\begin{tabular}{lcc}
\hline
Quasar   &   z & $logL_O$    \\
         &     & erg $s^{-1}$   \\
\hline
         &      Radio Quiet & \\
\hline
RXJ16331+4157 & 0.136  & 29.14  \\
MS02388-2314  & 0.284  & 29.92  \\
Q0335-350     & 0.321  & 29.68  \\
MS03363-2546  & 0.334  & 30.20  \\
US3333        & 0.354  & 30.55   \\
PHL6625$^a.$       & 0.38   & 29.98  \\
MS12186+7522  & 0.645  & 30.77  \\
Q1234+1217    & 0.664  & 30.67  \\
MS21340+0018  & 0.805  & 30.31   \\
SBS0954+495   & 1.687  & 31.32    \\
\hline
         &      Radio Loud   &    \\
\hline
3C219       & 0.174  & 29.93   \\
PKS1334-127 & 0.539  & 31.08   \\
3C207       & 0.684  & 30.98    \\
3C212       & 1.043  & 30.91   \\
S4 0917+624 & 1.446  & 31.29   \\
S4 0917+449 & 2.18   & 32.01   \\
4C71.07$^b$   & 2.19   & 33.04    \\
PKS2351-154 & 2.665  & 32.33   \\
PKS0438-43  & 2.852  & 32.46   \\
PKS0537-286 & 3.119  & 32.16   \\
S4 0636+680 & 3.174  & 33.04   \\
PKS2126-158 & 3.266  & 33.25   \\
PKS1442+101 & 3.53   & 32.7    \\
S41745+624  & 3.886  & 32.97   \\
\hline
\end{tabular}
\end{centering}

a.  4.6' from NGC247 nucleus
\newline
b. Blazar, EGRET $\gamma-$ray
\end{table}

The distribution of `candidate' cut-offs is striking.  Table 1
gives the total number of quasars, the number of ``candidate''
cut-off quasars and its fraction in each redshift bin, while
figure 8 shows the fraction of `candidate' cut-off quasars as a
function of z for both quasar samples.  The fraction of
``candidate'' cut-off quasars among the 18 radio-loud objects at
z$>2.2$ is significantly different from that at z$<2.2$
(probability of 0.002 \%, using the binomial distribution).  For
radio-quiet objects the absence of cut-offs among the 12 quasars
at z$>2.2$ is consistent with the cut-offs distribution at lower
redshift (probability of 50 \%).  In turn, the probability of
finding zero `candidate' cut-off quasars among the n=12 radio
quiet quasars at z$>2.2$, assuming a frequency of cut-offs
similar to that of radio loud quasars at the same redshift, is
0.3 \%.

This strongly suggests a difference in the `candidate' cut-off
quasar distribution with z between radio loud and radio quiet
objects.  We calculated, using the Fisher exact probability test
(e.g. Siegel 1956), the probability that the two ``candidate''
cut-off samples differ in the proportion with which they are
distributed in redshift, e.g. below and above a given
redshift. For z=2.2 the difference is significant at the 98.4\%
level.

\section{X-ray Spectra}

A flatter $\alpha_S$ at high redshift than at low redshift can be
due to at least three effects: (a) the redshifting of a soft
component contributing at the emission below $\approx 1-3$ keV
(quasar frame) out of the observed energy range for z$\gs2$; (b)
evolution of the emission spectrum; (c) a cut-off due to low
energy absorption becoming more frequent at z$\gs2$.
A simple color analysis cannot distinguish between these three
possibilities.  A two parameter spectral fit can discriminate
more strongly, and is possible for the quasars with a few hundreds
detected PSPC counts. This section investigates such fits.

Most of the quasars in the sample do not have published PSPC
spectral fits.
%
%
We extracted the full pulse height spectrum for each of the
quasars in the sample from the appropriate event files in the
ROSAT archive. We used standard extraction criteria (see
e.g. Fiore et al. 1994, Elvis et al 1994a).  Table 4 gives the
total counts in the 0.1-2.4~keV energy band, the exposure and the
off-axis angle for these quasars.  All the radio-loud quasars
were observed on-axis and were the target of their
observation. All the radio-quiet quasars (except RXJ16331+4157)
are instead serendipitous sources in ROSAT fields. With one
exception (SBS0945+495) the radio-quiet quasars all lay within the
inner PSPC rib ($r=18~$arcmin), where the calibration is most
accurate.

We then fitted the spectra with a power law model with low energy
absorption (at z=0). We made four fits for each quasar: We first
let the column density be free to vary, and then kept it fixed to
the Galactic (21 cm) value along the line of sight.  Similarly, we let
the power law spectral index be free to vary, and then kept it
fixed to the mean value of $\alpha_H$ at the redshift of the
quasar (Table 1).  The results are given in Tables 5 and 6 for
each quasar sample. Listed are: the best fit parameters, the
$\chi^2$, and the probability that the improvement in $\chi^2$
between the fits with free $N_H$ and $N_H$ fixed to the Galactic
value is significant (calculated using the $\chi^2$ distribution
with 1 dof for the $\Delta\chi^2$).

\begin{table}[ht]
\caption{\bf ``Candidate'' Cut-off Quasars: PSPC Observations}
\begin{tabular}{lccc}
\hline
Quasar        &  Counts & exposure & off-axis angle \\
              &         & sec      & arcmin \\
\hline
         &      Radio Quiet & &   \\
\hline
RXJ16331+4157 & 281  & 13636 & on \\
MS02388-2314  & 223  & 8008  & 16.4\\
Q0335-350     & 744  & 17957 & 13.0\\
MS03363-2546  & 938  & 50058 & 18.7 \\
US3333        & 416  & 11863 & 13.2\\
PHL6625       & 328  & 19072 & 3.0 \\
MS12186+7522  & 321  & 6484  & 14.6\\
Q1234+1217    & 114  & 9426  & 13.5 \\
MS21340+0018  & 86   & 5201  & 11.6 \\
SBS0954+495   & 182  & 3669  & 35 \\
\hline
         &      Radio Loud   &     &  \\
\hline
3C219       & 509   & 4386  & on \\
PKS1334-127 & 516   & 3614  & on \\
3C207       & 452   & 7017  & on \\
3C212       & 770   & 21565 & on \\
S4 0917+624 & 366   & 19465 & on \\
S4 0917+449 & 640   & 3367  & on \\
4C71.07     & 5254  & 6993  & on \\
PKS2351-154 & 419   & 6335  & on \\
PKS0438-43  & 163   & 21231 & on \\
PKS0537-286 & 555   & 9487  & on \\
S40636+680  &  68   & 5342  & on \\
PKS2126-158 & 1262  & 7392  & on \\
PKS1442+101 & 655   & 15433 & on  \\
S41745+624  & 506   & 16141 & on  \\
\hline
\end{tabular}
\end{table}

\begin{table}[ht]
\caption{\bf ``Candidate'' Cut-off Radio Quiet Quasars: Spectral
Fits}
\begin{tabular}{lccccc}
\hline
Quasar        &  $N_{HGal}$ & $N_H$ & $\alpha_E$ & $\chi^2$/dof
& prob.($\Delta\chi^2$ \\
              & $10^{20}$ cm$^{-2}$ & $10^{20}$ cm$^{-2}$ & & & \\
\hline
         &      Radio Quiet & & & &  \\
\hline
RXJ16331+4157 & 1.05 & $1.1^{+2.1}_{-1.1}$  & $0.87^{+0.80}_{-0.60}$
& 7.0/13 &  \\
              &      & $2.7^{+0.8}_{-0.4}$  & 1.6 FIXED      & 9.2/14  & -- \\
MS02388-2314  & 2.29 & $5.5^{+7.3}_{-3.1}$  & $1.0\pm0.8$    & 4.8/14  &    \\
              &      & 2.29 FIXED           & $0.32\pm0.25$  & 5.9/15  & -- \\
              &      & $8.2^{+3.8}_{-3.0}$  & 1.6 FIXED      & 5.9/15  & -- \\
Q0335-350     & 1.26 & $2.1^{+1.0}_{-0.8}$  & $1.18\pm0.35$  & 14.8/23 &    \\
              &      & 1.26 FIXED           & $0.86\pm0.08$  & 17.2/24 & -- \\
              &      & $3.1\pm0.3$          & 1.6 FIXED      & 17.6/24 & -- \\
MS03363-2546  & 1.05 & $2.1^{+1.8}_{-0.8}$  & $1.28\pm0.30$  & 72.6/67 &   \\
              &      & 1.05 FIXED           & $0.89\pm0.06$  & 77.4/68 &
97.2 \% \\
              &      & $2.9\pm0.2$          & 1.6 FIXED      & 74.3/68 &  \\
US3333        & 4.36 & $7.7^{+5.1}_{-2.5}$  & $1.36\pm0.55$  & 8.2/15  &   \\
              &      & 4.36 FIXED           & $0.77\pm0.14$  & 12.8/16 &
96.8 \% \\
              &      & $9.0\pm0.2$          & 1.6 FIXED      & 8.7/16
& -- \\
PHL6625       & 1.47 & $7.7^{+3.0}_{-2.8}$  & $1.88\pm0.6$   & 2.5/12 &    \\
              &      & 1.47 FIXED           & $0.32\pm0.14$  & 19.3/12
&  99.995 \% \\
MS12186+7522  & 3.02 & 3.0 FIXED            & $0.7\pm0.5$    & 5.7/11 & \\
              &      & $6.5^{+1.7}_{-1.0}$  & 1.6 FIXED      & 9.5/12 & -- \\
Q1234+1217    & 2.51 & $4.5^{+9.1}_{-3.1}$  & $1.5^{+1.6}_{-0.5}$ & 3.4/6
& \\
              &      & 2.51 FIXED           & $0.93\pm0.33$       & 3.8/7
& -- \\
MS21340+0018  & 4.00 & $2.7^{+18.3}_{-2.7}$ & $0.6^{+1.5}_{-0.3}$ & 0.71/5
&   \\
              &      & 4.0 FIXED            & $0.90\pm0.45$  & 0.81/6 & -- \\
              &      & $6.2^{+6.4}_{-1.7}$  & 1.6 FIXED      & 1.62/6 & -- \\
SBS0954+495   & 0.87 & $5.2^{+6.8}_{-5.2}$  & $2.0\pm2.0$    & 6.5/7  & \\
              &      & 0.87 FIXED           & $0.57\pm0.30$  & 7.9/8  & -- \\
              &      & $4.0^{+1.4}_{-1.0}$  & 1.6 FIXED      & 6.6/8  & -- \\
\hline
\end{tabular}
\end{table}

\begin{table}[ht]
\caption{\bf ``Candidate'' Cut-off Radio Loud Quasars: Spectral Fits}
\begin{tabular}{lccccc}
\hline
Quasar        &  $N_{HGal}$ & $N_H$ & $\alpha_E$ & $\chi^2$/dof
& prob.($\Delta\chi^2$ \\
              & $10^{20}$ cm$^{-2}$ & $10^{20}$ cm$^{-2}$ & & & \\
\hline
3C219       & 1.48 & $2.1^{+0.8}_{-1.2}$ & $0.22\pm0.40$ & 6.3/22  &  \\
            &      &  1.48 FIXED         & $0.05\pm0.10$ & 6.8/23  & -- \\
            &      & $5.7^{+0.7}_{-0.5}$ & 1.34 FIXED    & 23.0/23 & \\
PKS1334-127  & 4.41 & $7.1^{+3.0}_{-2.0}$ & $1.2\pm0.4$   & 11.0/20 &  \\
            &      & 4.41 FIXED          & $0.66\pm0.13$ & 15.3/21 & 96.2 \%\\
3C207       & 4.07 & $29^{+30}_{-21.5}$  & $2.3\pm1.4$   & 13.6/17 & \\
            &      & 4.07 FIXED          & $0.34\pm0.16$ & 24.9/18 & 99.92 \%\\
3C212       & 3.70 & $32^{+18}_{-16}$    & $1.9\pm1.0$   & 14.4/21 & \\
            &      & 3.70 FIXED          & $0.14\pm0.09$ & 32.0/22
& 99.997 \% \\
S4 0917+624 & 3.55 & $11.9^{+32}_{-7.9}$ & $0.46\pm0.30$ & 5.0/14  & \\
            &      & 3.55 FIXED          & $0.76\pm0.21$ & 8.25/15 & 92.6\\
S4 0917+449 & 1.51 & $2.9^{+1.2}_{-1.0}$ & $0.79\pm0.33$ & 17.6/22 & \\
            &      & 1.51 FIXED          & $0.37\pm0.08$ & 22.1/23
& 96.6\% \\
4C71.07     & 2.95 & $3.3\pm0.3$         & $0.52\pm0.09$ & 26.4/29 &  \\
            &      & 2.95 FIXED          & $0.43\pm0.03$ & 28.9/30 & -- \\
PKS2351-154 & 2.39 & $5.4^{+3.5}_{-1.2}$ & $0.87\pm0.49$ & 12.6/19 & \\
            &      & 2.39 FIXED          & $0.25\pm0.14$ & 17.3/20
& 97.0 \% \\
PKS0438-43  & 1.50 & $6.9^{+3.5}_{-1.8}$ & $0.70^{+0.27}_{-0.22}$
& 10.3/22 & \\
            &      & 1.5 FIXED           & $-0.16\pm0.06$ & 55.4/23
& $>99.999$ \\
PKS0537-286 & 2.06 & $2.7^{+1.7}_{-1.4}$ & $0.38\pm38$   &  16.2/22 & \\
            &      & 2.06 FIXED          & $0.22\pm0.11$ & 16.8/23  & -- \\
            &      & $3.8\pm0.5$         & 0.7 FIXED     & 17.8/23  & -- \\
S40636+680  & 5.7  & 5.7 FIXED           & $-0.1\pm0.4$  & 10.64/15 & \\
            &      & $20^{+10}_{-8}$     & 1.7 FIXED     & 9.52/15  & \\
PKS2126-158 & 4.85 & $12.9^{+7.2}_{-3.8}$ & $0.70^{+0.41}_{-0.29}$ & 20.6/20\\
            &      & 4.85 FIXED          & $-0.03\pm0.03$ & 49.56/20
& $>99.999$ \\
PKS1442+101 & 1.70 & $1.9^{+1.2}_{-0.9}$  & $0.46\pm0.35$ & 23.2/24 & \\
            &      & 1.70 FIXED          & $0.41\pm0.10$ & 23.4/23 &  --  \\
S41745+624  & 3.31 & $6.8^{+0.30}_{-3.0}$ & $0.78^{+1.0}_{-0.44}$ & 14.5/16 &\\
            &      & 3.31 FIXED          & $0.26\pm0.13$ & 19.1/17 &
96.8 \%\\
\hline
\end{tabular}
\end{table}

\bigskip
{ }From Tables 5,6 we see that the evidence of a cut-off is very
robust (`Class A') for four radio-loud quasars (PKS2126$-$158,
PKS0438$-$436, 3C~212, and 3C~207), Probability $\gs99.9 \%$.
For another four radio-loud quasars (S4~1745+624, PKS2351$-$154,
S4~0917+449 and PKS1334$-$127) the probability of a cut-off is
$\gs 95\%$ (`Class B').
For S4~0917+624 the probability for a cut-off is $\gs 92 \%$
(`Class C').
Furthermore three quasars (3C219, PKS0537-286 and S4~0636+680)
have very flat energy index, much flatter than the average for
radio loud quasars at those redshifts, if we insist on only
Galactic absorption (we also call these `Class C'). Excess
absorption, similar to that required in the better spectra,
readily produces a normal energy index.
Two radio-loud quasars do not show any evidence for a cut-off (4C71.07
and PKS1442+101).  The first is a famous blazar, detected by EGRET in
the GeV energies (e.g. von Montigny et al 1995) and therefore beaming
is important. The second is often classified as a Compact Steep Radio
Spectrum quasar (e.g. DallaCasa et al 1995).
The total number of cut-off radio loud quasars is therefore eight,
with three more objects likely to be cut-off quasars.

By contrast, in radio-quiet quasars we have a strong evidence for a
cut-off in only one case out of 289 (PHL6625, Probability $>99.9 \%$).
PHL6625 lies just 4.6 arcmin from the position of the low redshift
galaxy NGC~247. The cut-off in PHL6625 may well be due to absorption
in gas associated to NGC247 (Elvis et al., 1997).  Two other radio
quiet quasars MS03363-2546, US3333) have a probability for a cut-off
$\gs 95 \%$.  US3333 is included in the area of the sky already
surveyed by the NVSS (Condon et al., in preparation) but has not been
detected. The upper limit on the radio flux ($f_r < 2.5$ mJy) puts an
upper limit of 0.2 on $\alpha_{ro}$, very close to the threshold of
0.19 that we use to divide radio-loud and radio-quiet quasars. More
sensitive optical and radio observation are then needed to assess the
nature of this source.  MS03363-2546 is not detected in radio by
Stocke et al 1992. The limit on the radio flux assures that this
quasar is a truly radio-quiet source.  The origin of the cut-offs of
MS03363-2546 and US3333 is unknown and deserves follow-up studies.
The probability of finding by chance the detected number of cut-off
radio quiet quasar at z$<2.2$, assuming a frequency of cut-offs
similar to that of radio loud quasars is 1.3 \%.

\section{Comparison with previous results}

Results on the ROSAT pointed observations of the quasars in Table
3 have been published by Elvis et al (1994a, PKS0438-436,
PKS2126-158, S40636+680), Elvis et al (1994b 3C212), Maraschi et
al (1995, PKS1334-127), Brunner et al (1994, 4C71.07), Buhler et
al. (PKS0537-286), Bechtold et al.  (1994, PKS1442+101 and
S41745+624).

The spectral fitting results in Table 6 generally agree well with
the results presented in the above papers.  Maraschi et al (1995)
use a Galactic $N_H$ higher than the one we adopted (from Stark
et al 1992) by about $1.6\times10^{20}$ cm$^{-2}$, and as a
result conclude that there is little evidence for absorption in
this source.

Some of the quasars in Table 3 have also been observed by ASCA
(Siebert et al., 1996 PKS0537-286, Cappi et al., 1997,
PKS0438-436, PKS0537-286, 4C71.07, PKS2126-158).  The results
found by these authors are consistent with those obtained from
the ROSAT observations with the exception of 4C71.07, for which
Cappi et al. (1997) report significant intrinsic absorption (of about
$8\times10^{20}$ cm$^{-2}$). The comparison between the ROSAT and
the ASCA data implies a variation in the absorber column density
on a time-scale of less than 2.6 years in this source.
Cappi et al. (1997) discuss in detail the possibility that low energy
absorption in addition to the 21 cm value in their sample of high
redshift quasars may be due to molecular gas. For the four quasars in
common between their and our sample they find no strong evidence for
Galactic molecular gas absorption.  We note here that our limit on the
Galactic (21cm) $N_H$ effectively excludes low Galactic latitude
sources from our sample, thus minimizing the probability for a
contamination from Galactic molecular clouds, which are strongly
concentrated toward the Galactic plane. The CO survey by Blitz,
Magnani \& Mundy (1984) finds that molecular gas is uncommon at high
Galactic latitudes.  The only quasar known to be affected by molecular
gas absorption (NRAO140, Marscher 1988, Turner et al., 1995) lies at
low Galactic latitude, and was excluded from our sample because it has
a Galactic (21 cm) $N_H$ of $1.4\times10^{21}$ cm$^{-2}$.  Furthermore,
de Vries, Heithausen \& Thaddeus (1987) found that high Galactic
latitude molecular clouds differ from the Galactic plane clouds so
that their CO-to-H$_2$ conversion factor is significantly smaller:
$\sim 0.5\times 10{20}$ instead of
$\sim2-4\times10{20}$molecules~cm$^{-2}$(K$^{-1}$ km$^{-1}$s)$^{-1}$
(Combes, 1991).  This would reduce the additional hydrogen column
density implied by any CO emission by a factor $\sim$6.

The evolution of the quasar X-ray spectrum has also been studied
by three groups: Schartel et al (1996a) who used a sample drawn from
the RASS containing all quasars with
$M_V<-23$ and more than 80 RASS counts; Schartel et al (1996b)
used the Large Bright Quasar Sample (LBQS) quasars observed
during the RASS (using both detections and non-detections);
Puchnarewicz et al (1996) used AGNs identified in the ROSAT
International X-ray/Optical Survey (RIXOS).

All the radio-quiet objects in Schartel et al (1996a) are at
z$<0.5$, while radio-loud objects are detected up to z=2.5.  The
indices of the radio-quiet objects are fully consistent with
those of the WGACAT quasars.  The PSPC spectral indices of
radio-loud objects show a significant flattening above z$\sim1$,
which is interpreted by Schartel et al in terms of a selection
effect and/or a redshift effect, if quasar spectra are not simple
power laws (but rather have a concave shape).  In the WGACAT
quasars we see little change in both $\alpha_S$ and $\alpha_H$
from 0.1$<$z$<$2, most of the evolution being confined at
redshifts higher than $\sim2$.

On the other hand most of the LBQS objects studied by Schartel et
al (1996b) are radio-quiet quasars. Schartel et al (1996b) find a
marginal (2 sigma) evidence for a flattening of LBQS spectra at
z$\gs$1.5. This result is consistent with our findings (see Section
3.1.2 above) based on WGACAT quasars. In particular the
slopes reported by Schartel et al (1996b) agree well with those
in Table 1. Again, the study of the evolution of the spectrum of
radio-quiet quasars is limited by the fact that high redshift
radio-quiet quasars are faint and not easy to observe with ROSAT.

Puchnarewicz et al (1996) find a mean energy index significant flatter
than those obtained from the WGACAT quasars.  They also find no
evidence for a change in $\alpha_X$ with z.  The flat average slope
reported by Puchnarewicz et al could be due to the selection criteria
used to define the sample (a flux limit of $3\times 10^{-14}$ erg
cm$^{-2}$ s$^{-1}$ in the ROSAT 0.4-2 keV `hard band') which favors
the inclusion in the sample of flatter AGN.  Furthermore, the
correlation between the optical to X-ray index with the X-ray spectral
index (like the one found in Seyfert galaxies and low redshift
quasars, see eg. Walter \& Fink 1993, Fiore et al. 1995) can select
preferentially flat X-ray quasars when considering a strictly X-ray
selected sample (low optical to X-ray index).  We also note (from
their figure 3) that in their sample there are a few quasars with a
very flat spectral index, possibly due to absorption, which tend to
lower the average index.  These quasars have a red optical continuum,
strengthening the evidence for absorption in these cases.

Also the fraction of radio-loud AGN in the RIXOS is unknown. The
energy index Puchnarewicz et al (1996) find at 2$<$z$<$3 is
similar to the index we find in the same redshift interval for
radio-loud objects.  At low redshift the radio quiet population
should dominate in number (because of its much larger volume
density). However, we note that a similar X-ray flux limited
survey (Schartel et al 1996a) is completely dominated by radio
loud objects at z$\gs$0.5, since radio-loud objects are brighter
in X-rays than radio quiet (e.g Zamorani et al 1981, Green et al
1996).  Therefore the RIXOS indices, at least at high redshift,
could be dominated by the radio-loud population.

\section{Conclusions}

We have studied the 2 color X-ray spectra of a sample of about
167 radio-loud quasars and 286 `bona fide' radio-quiet quasars.
Radio-loud quasars cover the whole redshift space from 0.1 to 4
rather uniformly, while there are only three radio-quiet quasars at
z$>2.5$, against 12 radio-loud quasars at the same redshifts. Any
conclusion on radio-quiet quasars then apply to z$\ls2$ only.

Concerning the low energy cut-off we have established:

\begin {enumerate}

\item
Low energy X-ray cut-offs are more commonly (and perhaps
exclusively) associated with radio-loud quasars.  Detailed
spectral fits allow us to add, with some confidence, that
photoelectric absorption is a likely origin of the `low energy
cut-offs'.  The conclusion is that {\em low energy X-ray cut-offs
are associated with the quasars}, and not with intervening
systems, since those would affect radio-quiet and radio-loud
quasars equally.

\item
Among radio-loud quasars those at high redshift have a lower mean
$\alpha_S$ than those at low z (P=0.04\%), with many lying in the
X-ray `cut-off' zone. Detailed spectral analysis of all
candidate cut-off quasars show four robust cut-off detections
at redshift higher than 2.2.  The probability that the fraction
of cut-off quasars among the 18 objects at z$>2.2$ is similar
to that of radio-loud quasars at z$<2.2$ is very small, about 0.002
\% (using the binomial distribution).  This indicates that cut-offs
were more common in the past than they are now. I.e.  {\em the
X-ray cut-offs show evolution with cosmic epoch.}
%

\item
The degeneracy between redshift and luminosity found in
flux limited samples of quasars
was tested by using a partial correlation analysis.
We found that while $\alpha_S$ is truly anti-correlated with
redshift at the $99.9\%$ confidence level, in the case of
$\alpha_H$ the observed anti-correlation with redshift is mostly
due to a strong dependence on luminosity. Therefore, the cut-offs
are an evolutionary, not a luminosity, effect.

\bigskip
Concerning the emitted X-ray spectra of quasars we have
established:

\item
The distribution of $\alpha_S$ of radio-quiet quasars is
consistent with that of $\alpha_H$ for z$\ls2$.  This uniformity
suggests a single emission mechanism dominating the whole ROSAT
band (c.f. Laor et al., 1997) up to a redshift of about 2.  We
find a marginal evidence for a flattening of $\alpha_H$ (P=4.5
\%) going from z$<$1 to z=2, in agreement with previous studies
(Schartel et al 1996b).  This can be due to a selection effect
even if quasar X-ray spectra are simple power laws, because at
high redshift the steepest (and therefore faintest) sources would
not be detected.  However, it is well known that ROSAT PSPC 0.1-2
keV spectral indices of Seyfert 1 galaxies and low redshift
radio-quiet quasars are much steeper than those observed above 2
keV (e.g. Walter \& Fink 1993, Fiore et al 1994, Laor et al
1997).  If the spectrum of these AGN is made up of two distinct
components that are equal at a typical energy $E_0$, the
flattening of $\alpha_H$ at z$>1$ would suggest that $E_0$ lies
in the range 2-4 keV (quasar frame).

\item
Radio-loud quasars at z$<$2.2 show a `concave' spectrum
($\alpha_H < \alpha_S$ by $\sim$0.2).  Both indices are much
flatter than those of radio-quiet quasars.  This new result is in
line with the suggestion of Wilkes \& Elvis (1987) that the X-ray
spectrum of radio-loud quasars may be due to an additional
component above that seen in radio-quiet quasars. However, it may
also imply different processes at work in radio-loud and
radio-quiet sources (as recently suggested by Laor et al., 1997).

At z$\gs2$ the average soft and hard indices are similar and both
significatively smaller than at lower redshifts. This could be
due to the soft component of radio-loud quasars being completely
shifted out of the PSPC band at z$>2$.  Most z$>2$ radio-loud
quasars in our sample have flat radio spectra.  Padovani et
al. (1997) suggested that these quasars are analogs to LBL BL
Lacs, that is BL Lac objects with maximum energy emission in the
IR-Optical band. Their high energy radiation should then be
dominated by inverse Compton emission. At z$>2$ we are then
likely seeing pure inverse Compton emission. At lower redshift
the ROSAT PSPC band could sample a mixture of inverse Compton
emission, the tail of the Synchrotron component peaking in the
infrared and thermal emission from the hypothesized accretion
disk.

\end{enumerate}

The radio and optical properties of the quasars with low energy
X-ray cut-offs will be discussed in more detail in a companion
paper (Elvis et al. 1997).

\bigskip
This research has made us of the BROWSE program developed by the
ESA/EXOSAT Observatory, NASA/HEASARC, and the NASA/IPAC
Extragalactic Database (NED) which is operated by the Jet
Propulsion Laboratory, California Institute of Technology, under
contract with the National Aeronautics and Space Administration.



\newpage

\begin{figure}
\plotone{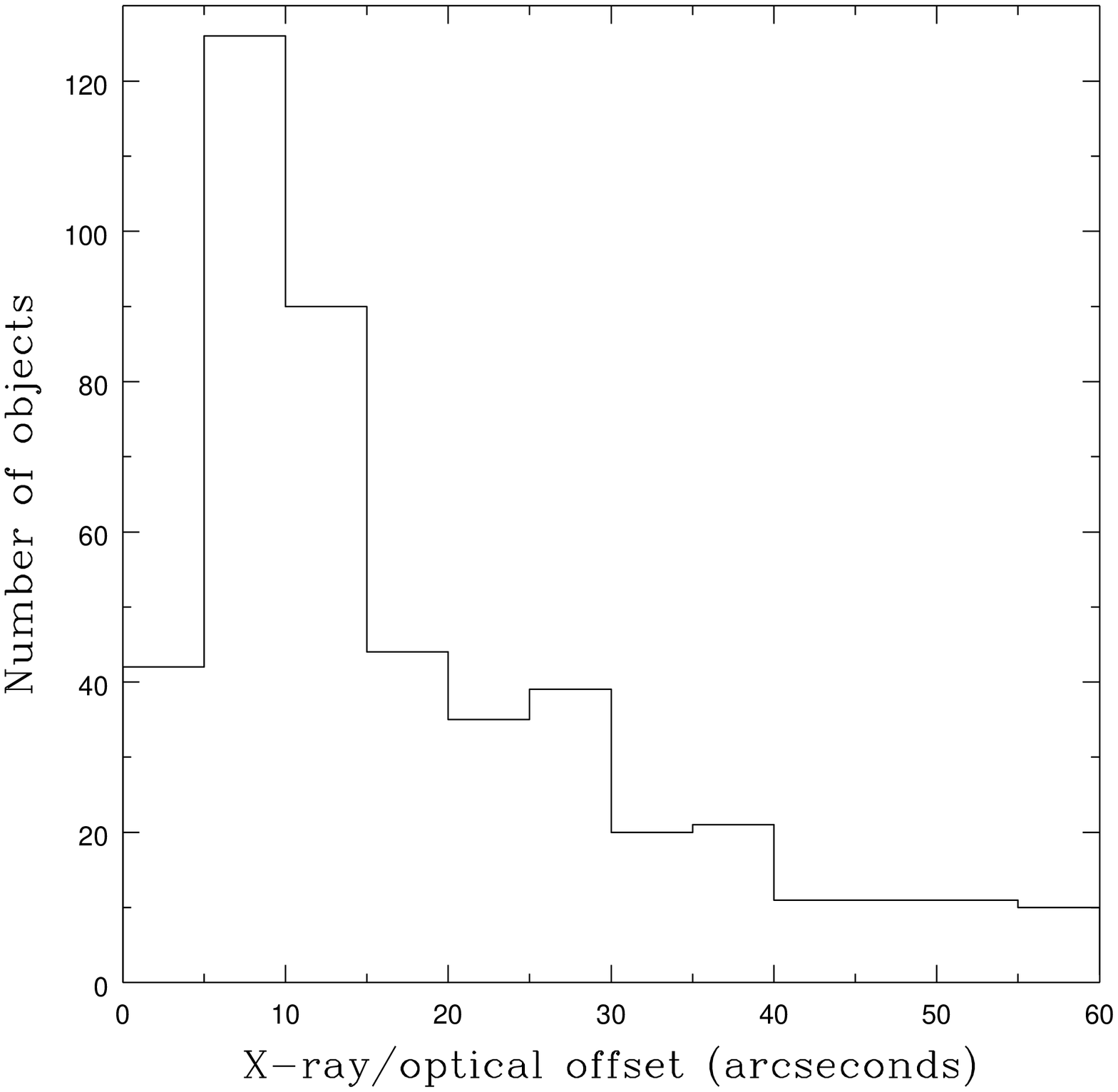}
\caption{The distribution of X-ray/optical offsets for our
sources, obtained by cross-correlating the WGA catalog with
various optical and radio catalogs with a correlation radius of
one arcminute. The mean offset is $\simeq 18$ arcsec. The number
of spurious associations is $\ls 2$ (see text for details).}
\end{figure}

\begin{figure}
\plotone{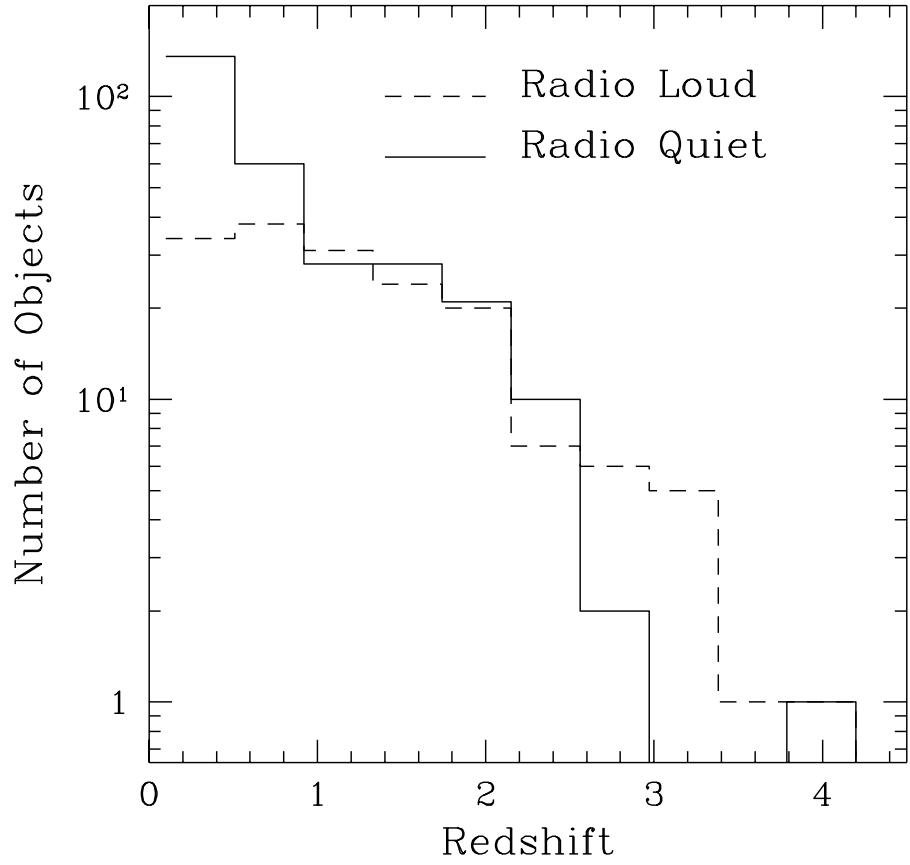}
\caption{The redshift distribution of WGACAT radio-quiet quasars
(solid line) and radio-loud quasars (dashed line), used in this
paper.}
\end{figure}

\begin{figure}
\plottwo{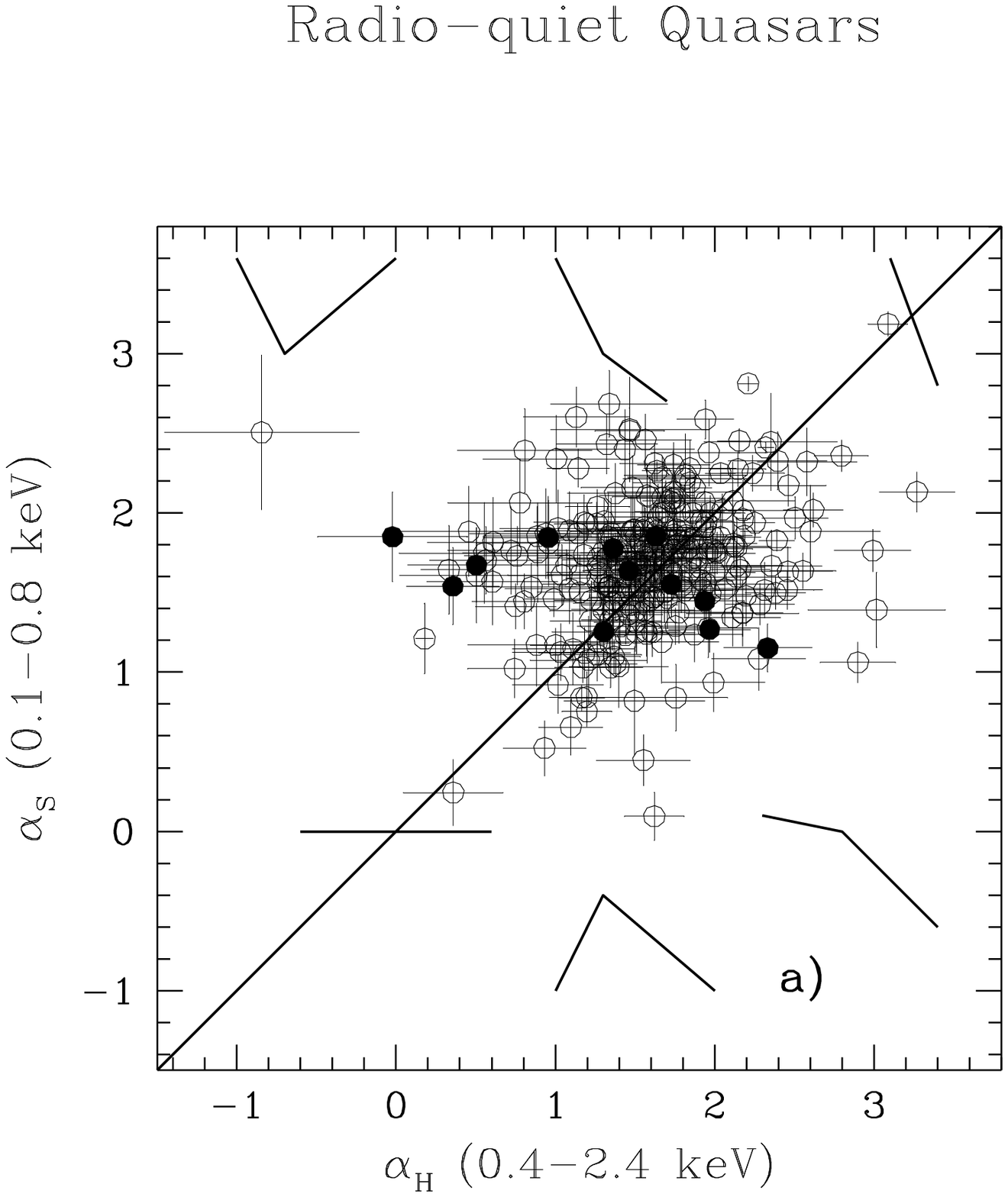}{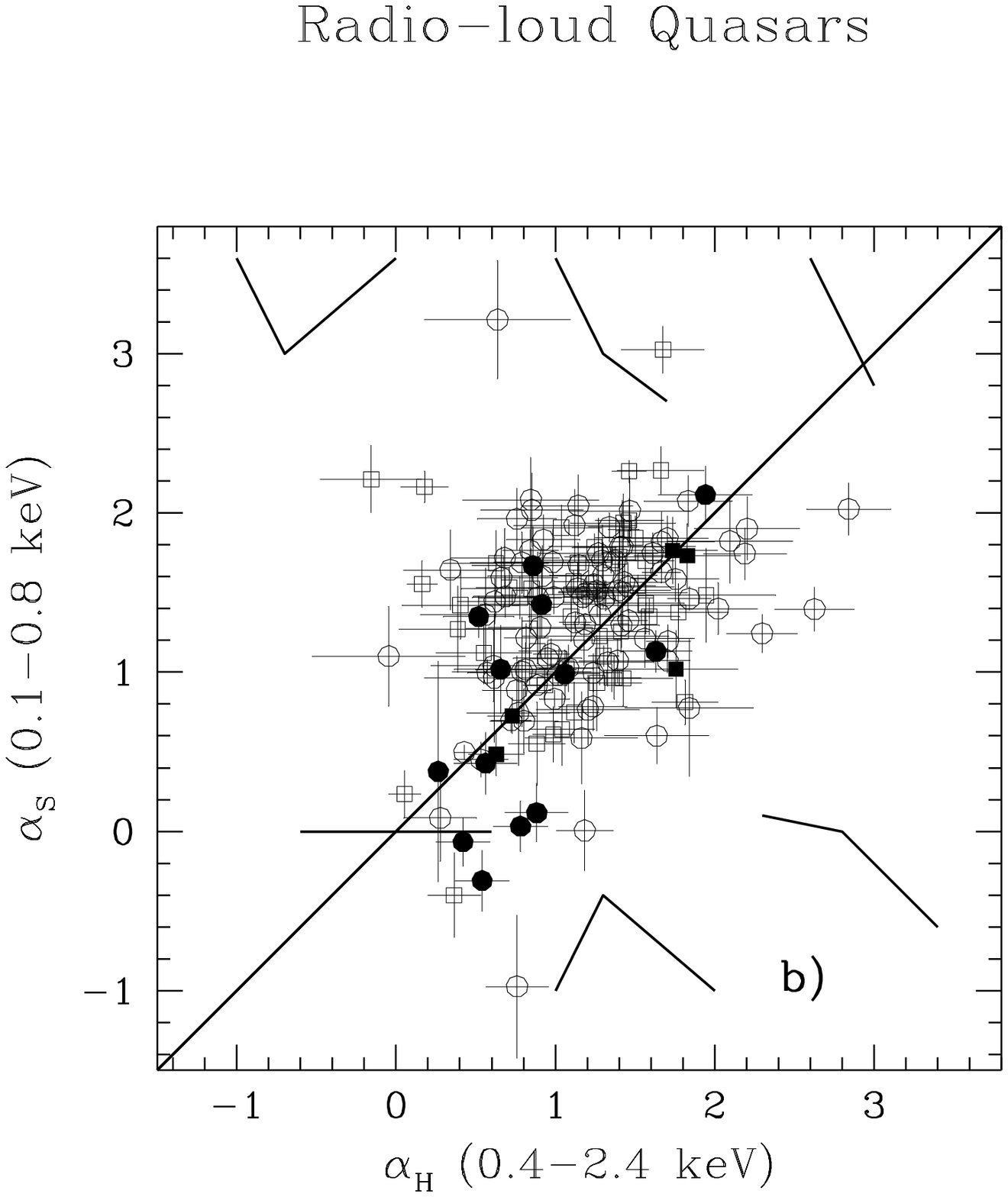}
\caption{ The `soft' effective spectral index ($\alpha_S$)
of (a) radio-loud quasars, and (b) radio-quiet quasars plotted
against the `hard' effective spectral index ($\alpha_H$).
Radio-loud flat spectrum radio sources are identified by circles,
steep radio spectrum sources by squares. High redshift quasars
(z$>$2.2) are shown with filled symbols.  Three-point spectra
illustrate the radically different spectral shapes in different
parts of the diagrams.  }
\end{figure}

\begin{figure}
\plotone{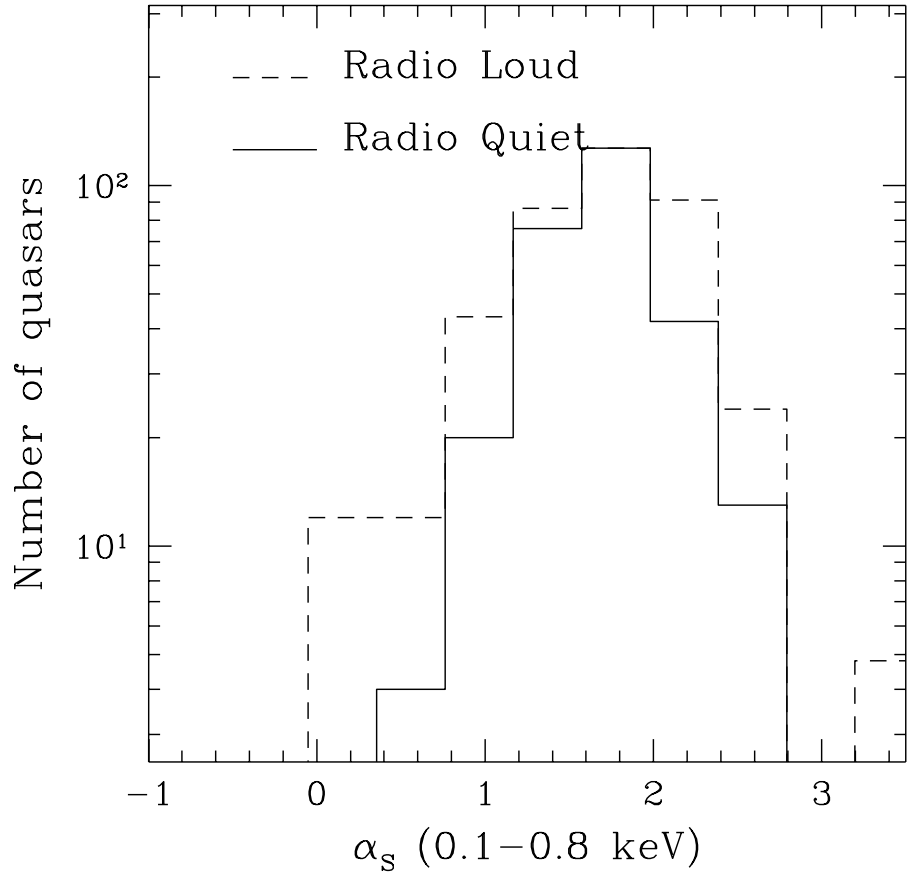}
\caption{ The distribution of radio-loud (dashed line) and
radio-quiet (solid line) indices about their respective mean
$\alpha_S$. The mean of radio-loud quasars has been offset to
coincide at $\alpha_S=1.7$ with that of radio-quiet quasars.  The
radio-loud distribution has been normalized to the maximum of the
radio-quiet distribution.}
\end{figure}

\begin{figure}
\plottwo{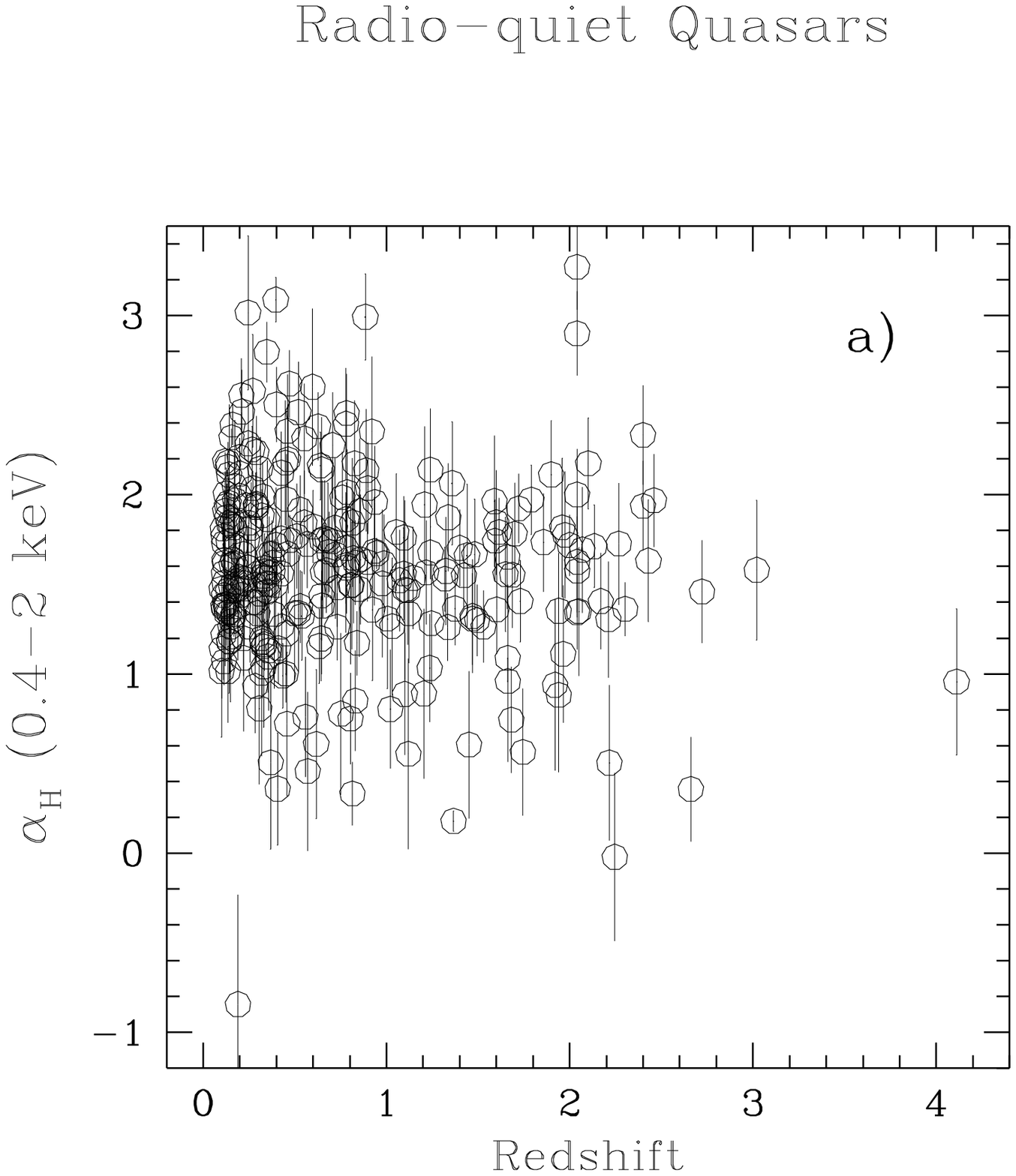}{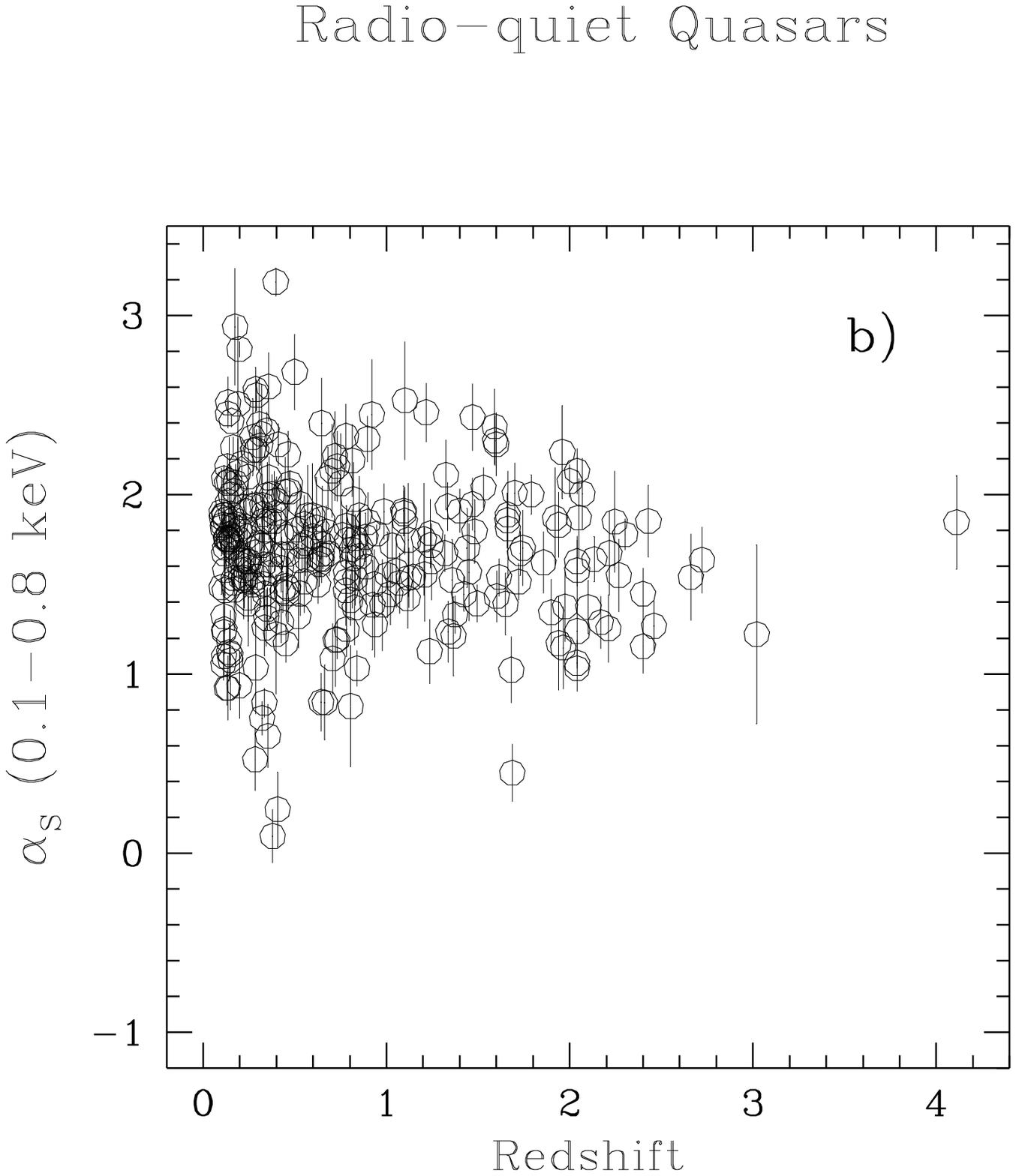}
\caption{$\alpha_H$ (a) and $\alpha_S$ (b) plotted against
the redshift for radio-quiet quasars. }
\end{figure}

\begin{figure}
\plottwo{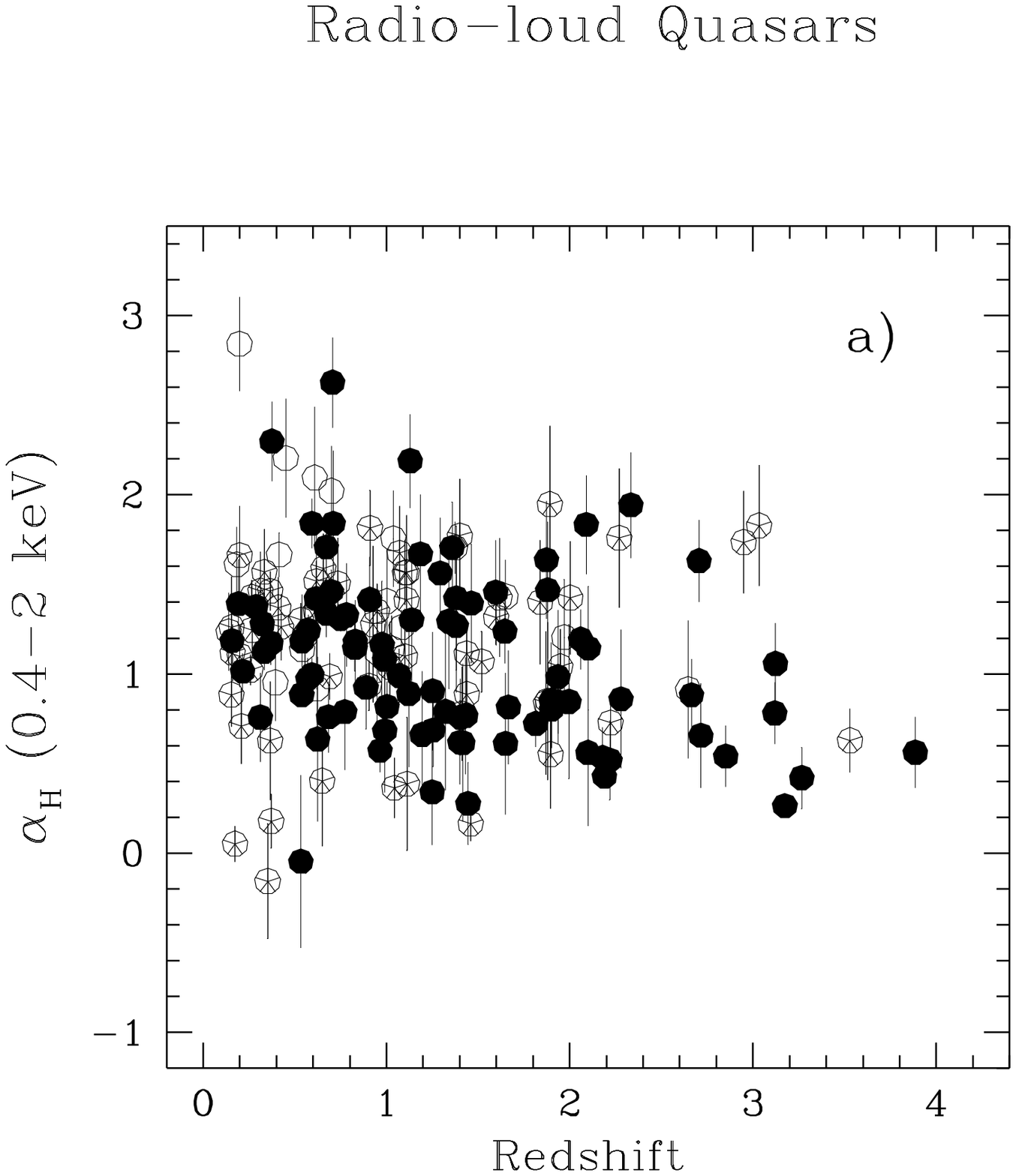}{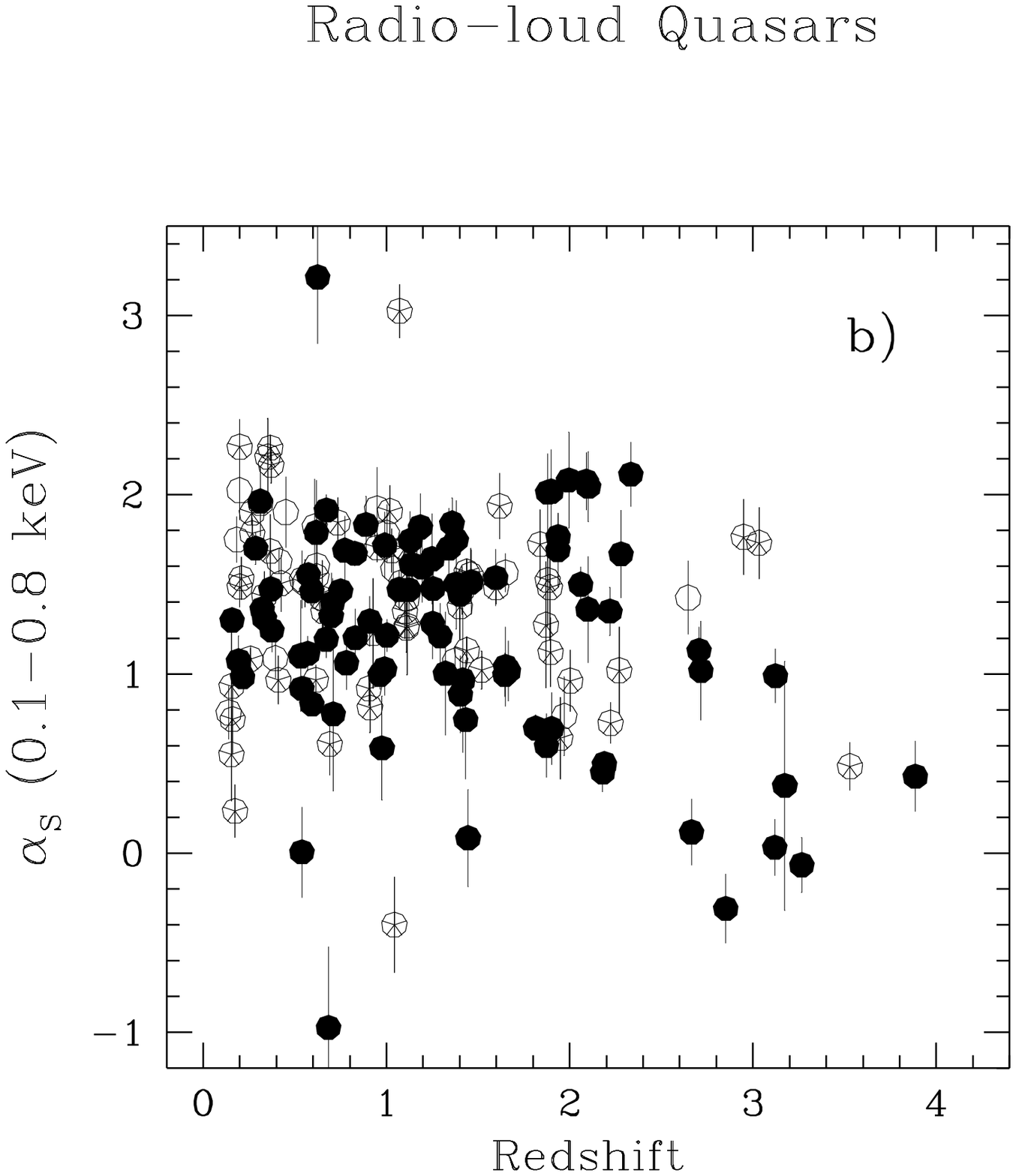}
\caption{ $\alpha_H$ (a) and $\alpha_S$ (b) plotted against the
redshift for radio-loud quasars. Filled circles identify flat
radio spectrum quasars. Starred circles identify steep radio
spectrum quasars. }
\end{figure}

\begin{figure}
\plotone{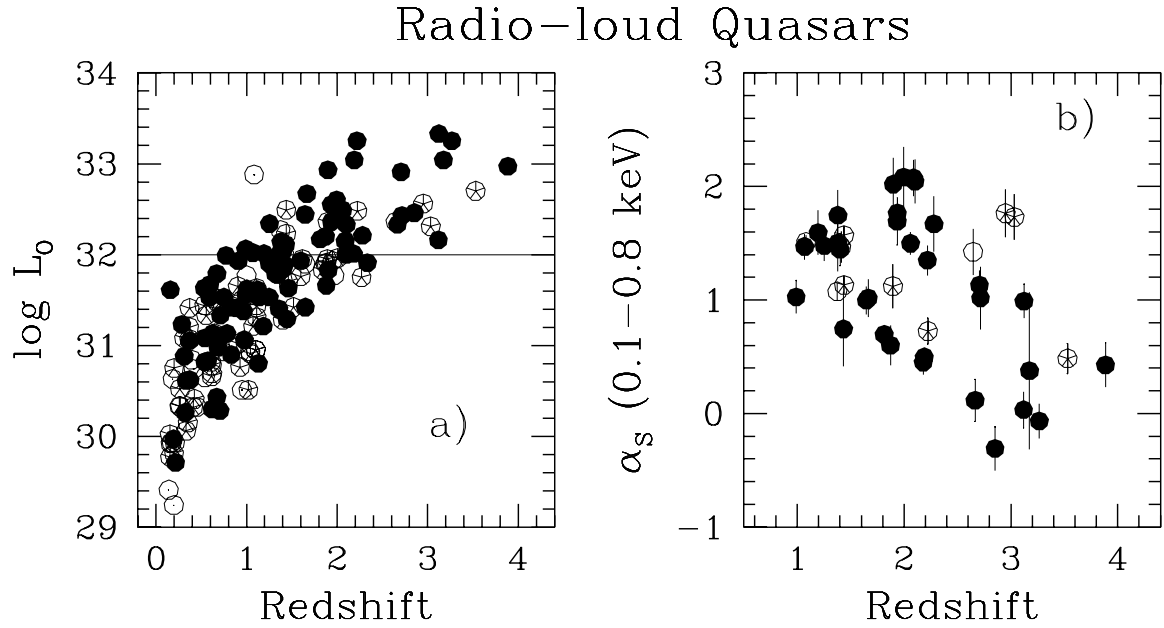}
\caption{(a) The redshift--optical luminosity correlation
for radio loud quasars.  (b) The soft spectral index
$\alpha_S$--redshift correlation for radio-loud quasars with high
optical luminosity ($>10^{32}$ erg s$^{-1}$. }
\end{figure}

\begin{figure}
\plotone{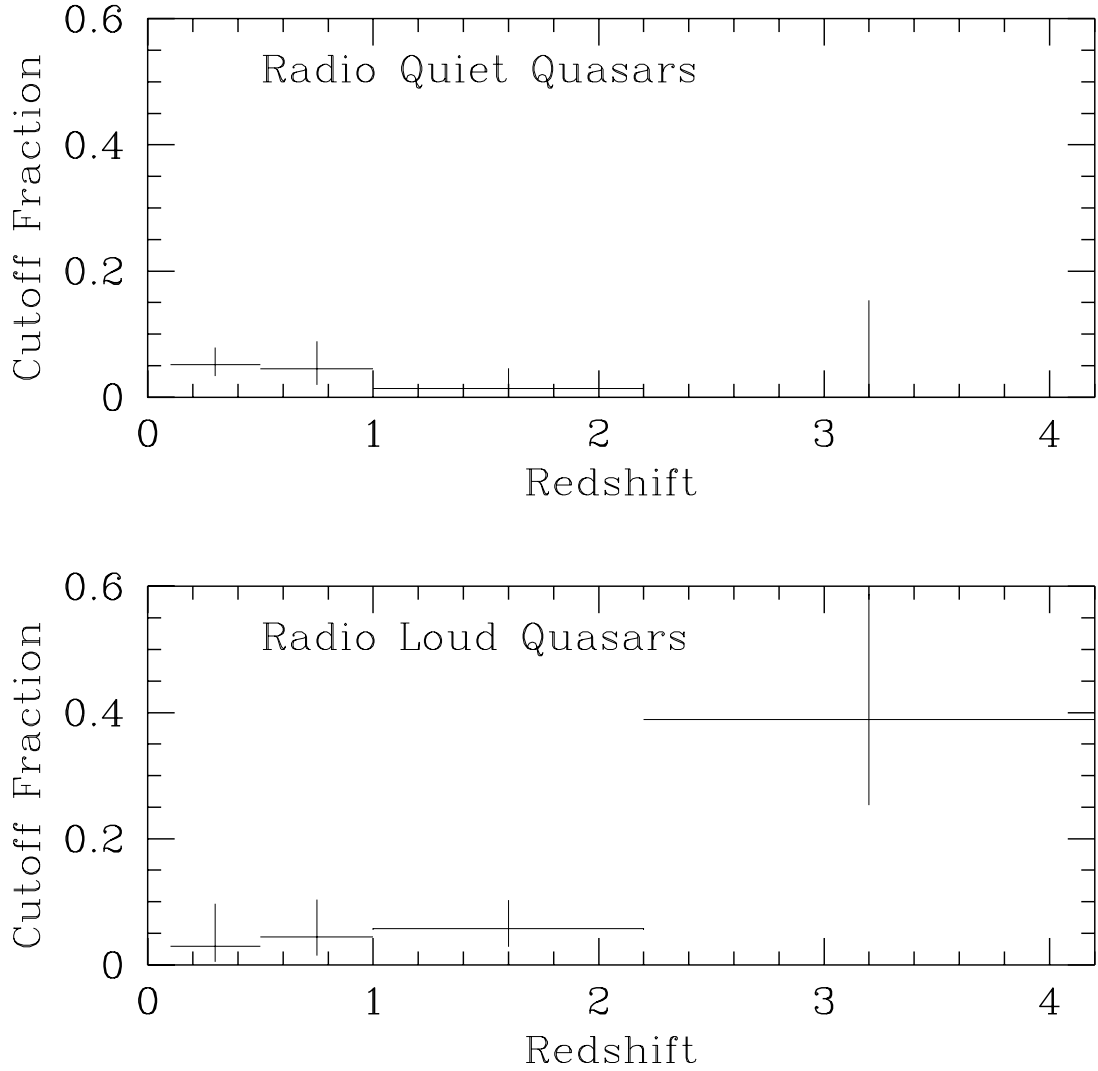}
\caption{ The fraction of ``candidate'' cut-off quasars as a
function of the redshift for radio quiet quasars (upper panel)
and radio-loud quasar (bottom panel). }
\end{figure}

\end {document}